\documentstyle[12pt]{article}
\begin{document}
\vspace{7mm}
\begin{center}

{\Large\bf Square-root quantization: application to quantum black holes}
\vspace{12mm}

{\large Victor Berezin}
\vspace{3mm} 

Institute for Nuclear Research of the Russian Academy of Sciences,
60th October Anniversary Prospect, 7a, 117312, Moscow, Russia
e-mail: berezin@ms2.inr.ac.ru

\end{center}
\vspace{14mm}
{\large\bf Introduction}
\vspace{3mm}

In this short paper we present two different quantizations of the simple
one-dimensional model - selfgravitating spherically symmetric thin shell. 
Fortunately enough, both ways of quantization lead to exactly solvable 
problems and we obtained mass spectra of the shell, analyzed them and 
extracted the quantum black hole spectra. 

The first approach can be called a proper time quantization. By using it 
we obtain a finite difference Schroedinger equation. Its detailed description, 
the method of finding the general solution, the needed boundary conditions, 
the discrete spectrum of eigenvalues in the case of bound motion and the 
resulting quantum black hoe mass spectrum were already published in \cite{me}.
Here we only pointed out the important steps that lead us to the discrete 
mass spectrum of quantum black holes.

Another approach can be called a Lorentzian time quantization. We show 
that there exists a suitable canonical transformation converting a famous 
square-root operator into an exponential operator. The resulting Schroedinger 
equation is, again, a finite difference equation of the same type as in the 
first approach. Then, repeating the steps we made in the first part of the 
paper we jump to the mass spectra of the shells and quantum black holes. 
The latter is similar to that found earlier.
\vspace{10mm}
{\large\bf Part One}
\vspace{3mm}
As was shown in \cite{us}, the classical evolution of a selfgravitating 
spherically symmetric thin dust shell is described completely by the single 
equation

\begin{equation}
m = M \sqrt{\dot\rho^2 + 1} - \frac{GM^2}{2\rho}
\label{clas}
\end{equation}

where $m$ is a total mass (energy) of the system, $M$ is a bare mass of the 
shell, $\rho(\tau)$ is its radius as a function of the shell's proper time, 
and $G$ is the gravitational constant.

Using Eqn.(\ref{clas}) as a pre-Hamiltonian we can introduce a momentum $p$ 
corresponding to the proper time velocity $\dot\rho$ and calculate the 
Hamiltonian

\begin{equation}
H = M \left( \cosh{p} - \frac{\kappa M^2}{2x}\right) .
\label{newH}
\end{equation}
where $x=M\rho$ is a dimensional radius. Such a Hamiltonian leads to the 
following Schroedinger equation in finite differences

\begin{equation}
\Psi(x+i) + \Psi(x-i) = \left(2\epsilon + \frac{\alpha}{x}\right) \Psi(x) ~,
\label{mefinal}
\end{equation}
where $\epsilon = m/M$, $\alpha=GM^2$.

The general solution to this equation in the momentum representation is

\begin{eqnarray}
\Psi_p = C \frac{z}{(z-z_0)(z-\bar{z}_0)}
\left(\frac{z-\bar{z}_0}{z-z_0}\right)^{\beta},\nonumber \\
\nonumber\\
z = e^p,\  z_0 = e^{i\lambda},\  \bar{z}_0 = e^{-i\lambda},\nonumber \\
\epsilon = \cos\lambda,\  \alpha = GM^2 =  2\beta\sin\lambda.
\label{p}
\end{eqnarray}

The general solution in the coordinate representation can be written in the 
form

\begin{eqnarray}
\Psi_{gen} &=& \Psi_\beta(x) \sum_{k=-\infty}^{\infty} c_k e^{-2\pi x}, \\
\nonumber\\
\Psi_\beta &=& (-4\pi\beta e^{-i\lambda}\sin{\lambda}) x e^{-\lambda x}
\nonumber\\ 
&&F(1-ix, 1-\beta; 2; 1-e{-2i\lambda})
\end{eqnarray}
\noindent
where $F(a,b,c;z)$ is a Gauss's hypergeometric function.

Due to the lack of space we will not discuss here the problem of boundary 
condition (for the details see \cite{me}). The main goal of the present paper 
is to find a discrete mass spectrum of the bound states, and the desired 
spectrum comes not from the boundary conditions but from the analytical 
properties of the solutions. The result is

\begin{eqnarray}
\beta = n, \qquad n = 1, 2, ...\nonumber \\
\epsilon = \sqrt{1 - \frac{\alpha^2}{4 n^2}}, \nonumber \\
m = M \sqrt{1 - \frac{G^2M^4}{4 n^2}}.    
\label{betan}
\end{eqnarray}

It was argued in \cite{me} that the above expression is a mass spectrum of the 
selfgravitating shells but not that of the black holes. To find the latter one 
we should look at the function $m(M)$ more attentively. This function has two 
branches, the increasing and decreasing ones. The increasing branch describes 
the shells that do not collapse. The decreasing branch corresponds to the 
wormhole states. And only maxima of the curves $m(M)$ for fixed quantum 
numbers $n$ give us the black hole states. Thus, we have for the black hole 
mass spectrum   

\begin{equation}
m_{BH} = \frac{2}{\stackrel{4}{\sqrt{27}}}\frac{\sqrt{n}}{G} 
\approx 0.9 \sqrt{n} M_{Pl} .
\label{BHm}
\end{equation}

We will not discuss this formula here noting only that the same functional 
form was obtained by many authors, and J.Bekenstein was the first one \cite{J}.

\vspace{10mm}
{\large\bf Part Two}
\vspace{3mm}

Let us now return to the second way of dealing with square-root kinetic terms.

If we will use the Lorentz time velocity $v$ instead of the proper time 
velocity $\dot{\rho}$ the pre-Hamiltonian for our shell take the more familiar 
form

\begin{equation}
m = \frac{M}{ \sqrt{1 - v^2}} - \frac{GM^2}{2\rho}
\label{clastwo}
\end{equation}

Introducing now a momentum $\Pi$ corresponding to the Lorentz time velocity 
$v$ we arrive at the famous square-root Hamiltonian for the radial motion 
in the field of Coulomb potential

\begin{equation}
H = \sqrt{\Pi^2 + M^2} - \frac{\alpha}{2\rho}
\label{hnew}
\end{equation}
The quantum analogs of the radius and the conjugate momentum are operators 
subject to the well known commutation relations. The kinetic part of the 
Hamiltonian is a nonlocal operator. To reveal this non-locality more 
explicitly we make the following canonical transformation before going to the 
quantization procedure.

\begin{equation}
\Pi = M\sinh p , \qquad  \rho = \frac{y}{M\cosh p}  
\label{canon}
\end{equation} 

The Hamiltonian \ref{hnew} now becomes
\begin{equation}
H(y,p) = M \cosh p (1 - \frac{\alpha}{2y}).
\label{hmy}
\end{equation}

The expression for the quantum counterpart depends on the chosen operator 
ordering. We will use the Hamiltonian \ref{hmy} write the corresponding 
Schroedinger equation in the form

\begin{eqnarray}
(y - \frac{\alpha}{2})\left(\Phi(y+i) + \Phi(y-i)\right) = 
2\epsilon y \Phi(y) ~,\nonumber \\
\nonumber\\
\Phi = (1 - \frac{\alpha}{2y})\Psi .
\label{mephi}
\end{eqnarray}

If we would choose the reverse ordering of momentum and coordinate functions 
we would get the same equation for $\Psi(y)$ instead of $\Phi(y)$ . We will 
not discuss here the important question of hermiticity of the quantum 
Hamiltonian.

The solution to Eqn.(\ref{mephi}) in the momentum representation is

\begin{eqnarray}
\Phi_p = C \frac{z^{1+i\frac{\alpha}{2}}}{(z-z_0)(z-\bar{z}_0)}
\left(\frac{z-\bar{z}_0}{z-z_0}\right)^{\beta},\nonumber \\
\nonumber\\
z = e^p,\  z_0 = e^{i\lambda},\  \bar{z}_0 = e^{-i\lambda},\nonumber \\
\epsilon = \cos\lambda,\  \alpha = GM^2,\ \beta = \frac{\alpha}{2}\cot\lambda.
\label{pnew}
\end{eqnarray}
This expression differs from the solution \ref{p} only by the factor 
$z^{i\alpha /2}$ . But if we shift the argument in the corresponding solution 
in the coordinate representation $y\to (y-\alpha/2)$ then the Fourier 
transform of this shifted function $\tilde\Phi_p$ will be exactly the same as 
$\Psi_p$ in Part One (Eqn.(\ref{p}). This means that the discrete spectrum 
in our second approach is determined by 

\begin{eqnarray}
\beta = n, \qquad n = 1, 2, ...\nonumber \\
m = \frac{M}{\sqrt{1 + \frac{G^2M^4}{4 n^2}}}.    
\label{betanew}
\end{eqnarray}

This is the so called Sommerfeld spectrum which was obtained in \cite{Haj} for 
the same classical model but using the Klein-Gordon Hamiltonian.

By the same procedure as in Part One we obtain the following black hole mass spectrum

\begin{equation}
m_{BH} = \sqrt{n} M_{Pl}.
\label{bhnew}
\end{equation}

The author is grateful to Russian Basic Research Foundation for essential 
financial support (Grant No 95-02-04911a).

\end{document}